\documentclass[twocolumns,traditabstract]{aa}
\usepackage{graphicx}
\usepackage{txfonts}
\def\co{CO(1-0)}
\def\cod{CO(2-1)}
\def\cot{CO(3-2)}

\begin{document}
   \title{\object{NGC6240}: extended CO structures and their association with shocked gas
   \thanks{
This work is based on observations carried out with the IRAM Plateau de Bure Interferometer. IRAM is supported by INSU/CNRS (France), MPG (Germany) and IGN (Spain). This work is also based on observations performed with the Chandra X-ray Observatory.
}
}

   \author{C. Feruglio \inst{1}
          \and
	F. Fiore  \inst{2} 
        \and
          R. Maiolino \inst{3}
         \and
         	E. Piconcelli  \inst{4}
	\and 
        H.  Aussel \inst{5}
        \and
      D. Elbaz \inst{5}
        \and
        E. Le Floc'h \inst{5}
	\and
         E. Sturm \inst{6}
        \and
        R. Davies \inst{6}
        \and
        C. Cicone \inst{3}
          }

   \institute{IRAM - Institut de RadioAstronomie Millim\'etrique 
300 rue de la Piscine, Domaine Universitaire 
38406 Saint Martin d'H\'eres, France, \email{feruglio@iram.fr}
         \and
            INAF- Osservatorio astronomico di Roma, via Frascati 33, 00040 Monteporzio Catone, Italy
            \and
            Cavendish Laboratory, University of Cambridge, 19 J. J. Thomson Ave., Cambridge CB3 0HE, UK
            \and
            XMM-Newton Science Operations Centre, ESAC, P.O. Box 78, 28691 Villanueva de la Can\~{a}da (Madrid), Spain
          \and
            Laboratoire AIM-Paris-Saclay, CEA/DSM/Irfu, CNRS, Universit\'e Paris Diderot, Saclay, pt courrier 131, 91191 Gif-sur-Yvette, France
            \and
           Max-Planck-Institut fur Extraterrestrische Physik (MPE), Giessenbachstr. 1, 85748 Garching, Germany
             }

\date{Received 4 June , 2012}

 
  \abstract{
We present deep \co~ observations of \object{NGC6240} performed with the IRAM
Plateau de Bure Interferometer (PdBI).
\object{NGC6240} is the prototypical example of a major galaxy merger in
progress, caught at an early stage, with an
extended, strongly-disturbed butterfly-like morphology and the presence
of a heavily obscured active nucleus in the core of each progenitor
galaxy.
The  CO line shows a skewed profile with very broad and asymmetric
wings detected out to velocities  of $-600$ km/s and $+800$ km/s 
with respect to the systemic velocity.
The PdBI maps reveal the existence of  two
prominent structures of blueshifted CO emission. 
One extends eastward, i.e. approximately perpendicular to the line connecting the galactic nuclei,
over scales of $\sim$7 kpc and shows velocities up to $-$400 km/s. 
The other extends southwestward out to $\rm \sim7~ kpc$ from the nuclear region, and has a velocity of $-$100 km/s 
with respect to the systemic one.  
Interestingly, redshifted emission with velocities 400 to 800 km/s is detected around the two nuclei, 
extending in the east-west direction, and partly overlapping with the
eastern blue-shifted structure, although tracing a more compact region 
 of size $\sim 1.7$ kpc.
The overlap between the southwestern CO blob and the dust lanes seen in
HST images, which are  interpreted as  tidal tails,
indicates that the molecular gas is deeply affected by galaxy interactions.
The eastern blueshifted CO emission is co-spatial with an
H$\alpha$ filament that is associated with strong H$_2$ and soft X-ray
emission. The analysis of Chandra  X-ray data 
provides strong evidence for shocked gas  at the  position of the
H$\alpha$ emission.
Its  association with outflowing molecular gas supports a scenario
where  the molecular gas is compressed into a shock wave that
propagates eastward from the nuclei. If this is an outflow, the AGN are likely
the driving force.
}


\titlerunning{Extended CO in \object{NGC6240}}
\authorrunning{C. Feruglio et al. }
 \maketitle

\section{Introduction}

The observed transformation of gas-rich  
star-forming galaxies into red, bulge-dominated spheroids devoid of gas,
is due to several mechanisms. 
 In massive galaxies star formation might lead to a faster gas consumption rate,
compared to less massive ones (Daddi et al. 2007, Peng et al. 2010,
Elbaz et al. 2011, Rodighiero et al. 2011). 
In addition,  galaxy
interactions,  mergers (Sanders et al. 1988, Barnes \& Hernquist 1996, Cavaliere
\& Vittorini 2000, Di Matteo et al. 2005), together with active galactic nuclei (AGN) and starburst
feedback are expected to play a role (Silk \& Rees 1999, King 2010 and references therein).
Mergers can destabilize cold gas and trigger both star formation and nuclear
accretion onto super-massive black holes (SMBHs), inducing AGN activity.
A natural expectation of this scenario is that the early,
powerful AGN phase is highly obscured by large columns of gas and dust (e.g. Fabian 1999).
Once a SMBH reaches masses $>10^{7-8}$ M$_\odot$, the AGN can efficiently 
contribute to the radiative heating of the inter stellar medium (ISM) through winds and shocks, 
thus inhibiting further accretion and also star-formation in the nuclear region 
and possibly at larger scales in the galactic disk.
The radiative feedback from a luminous AGN is therefore a 
mechanism that could explain the low gas content of local massive galaxies 
and the galaxy bimodal color distribution 
(Kauffmann et al. 2003, Croton et al. 2006, Menci et al. 2006). 
This evolutionary scenario needs to be observationally  confirmed. 
This can be achieved by observing systems during a major interaction
phase, that probe both AGN and starburst-driven winds, and their interaction
with the molecular gas, which represents the bulk of the gas in a galaxy.

Only recently molecular gas outflows have been
discovered in both star-forming galaxies (e.g. M82, Walter et
al. 2002, Arp 220, Sakamoto et al 2009) and in the hosts of powerful AGN, 
through the detection of both molecular absorption lines with P-Cygni profiles, and of broad 
molecular emission lines (Feruglio et al. 2010, Fisher et al 2010, Alatalo et al. 2011, Sturm et al. 2011, 
Aalto et al. 2012, Cicone et al. 2012, Maiolino et al. 2012).  
The  inferred outflow rates show that these 
outflows can displace large amounts (several hundreds of solar masses per year) of molecular gas into the galactic disk, 
hence supporting AGN feedback model predictions (e.g. King 2005, 2010, Zubovas \& King 2012, Lapi et al. 2005, Menci et
al. 2008). 
In particular, strong molecular outflows have been found in several local Ultra Luminous Infrared Galaxies (ULIRGs), 
suggesting that they might be common in objects undergoing major mergers (Sturm et al. 2011). 
The "prototype" of this class of objects is Mrk 231, in which 
we indeed discovered a massive molecular outflow extended on scales of $\sim$1 kpc 
in the host galaxy disk (Feruglio et al. 2010). 
Mrk231 is known to be in a late merger
state (Sanders et al. 1988, Davies et al. 2005), and shows a compact molecular disk
(Carilli et al. 1998).

In the framework of the exploration of massive molecular outflows in nearby ULIRGs and LIRGs, 
we present  in this work our millimeter observations of the nearby merger \object{NGC6240}. 
This is a prototypical galaxy undergoing transformations. 
Thanks to its close distance ($z$=0.024),  this system offers the 
opportunity to investigate in detail the distribution and
dynamics of the molecular gas during a merger event, which represents
the key process in hierarchical models of galaxy formation and
evolution. It is a massive object, resulting from the merger of two gas rich
spirals. 
The nuclei, separated by $\sim2$\arcsec in approximately the north-south direction, 
are located in the central region of the system, probably the remnants
of the bulges of the progenitor galaxies, since the majority of the nuclear
stellar luminosity is provided by stars predating the merger (Engel et
al. 2010).  Each nucleus hosts an AGN (Komossa et al. 2003). 
At least one of the AGN is highly obscured by a hydrogen column density of 
N$_H>10^{24}$ cm$^{-2}$ (Compton-thick), and has an intrinsic luminosity L(2-10 keV)$>10^{44}$ erg s$^{-1}$ (Vignati et
al. 1999). The mass of the SMBH powering this AGN likely exceeds
$10^8$ M$_{\odot}$ (Engel et al. 2010). 
The system is in an early, short-lived phase of merging, likely between the
first encounter and the final coalescence, as witnessed by intense star-formation
activity (see e.g. Sanders \& Mirabel 1996, Mihos \& Hernquist
1996). The system thus has a greater physical size with respect to
Mrk231, and exhibits large scale streamers and outflows, witnessed by the
spectacular butterfly-shaped emission-line nebula seen in HST
H$\alpha$ images (Gerssen et al. 2004). 
The nebula is interpreted as
evidence of a super-wind shock-heating the ambient ISM. The
emission line filaments and bubbles appear to trace a bipolar outflow pattern,
aligned east-westward, extending up to 15-20 arcsec (7-10 kpc) from the
nuclear region perpendicular to the wide dust lane seen in the HST images (Gerssen et al. 2004), and to the line
connecting the two nuclei. 
The superwind is likely powered
by both the nuclear star-formation and by the AGN. 
\object{NGC6240} is thus an ideal target to  study: a) the interplay
between AGN and star-formation activity; b) the mechanism of
transport of energy from the nuclei to the gas in the outer parts of the galaxy; 
c) how the molecular gas is heated by the winds.

We present in this work \co~ maps obtained with the IRAM Plateau de Bure Interferometer (PdBI)
 in the D and A array configurations.  
These data have lower spatial resolution than previous works (Engel et
al. (2010), Iono et al.( 2007), Nakanishi et al. (2005)), but the useful bandwidth is much
broader and the noise level is a factor $>2$ lower. 
We also present a reanalysis of the Chandra X-ray,
high spatial resolution data (available from the Chandra public archive). 
A $\Lambda$CDM cosmology ($H_0=70$ km s$^{-1}$ Mpc$^{-1}$; $\Omega_M$=0.3; $\Omega_{\Lambda}=0.7$) is adopted.

\section{PdBI observations and data analysis}

We observed with  the PdBI the \co~ transition, redshifted to 112.516 GHz assuming a systemic velocity of 7339 km/s (Iono et al. 2007), corresponding to a redshift of  z$ = $0.02448.  
The observations were carried out
in May 2011 with six antennas using the compact array 
configuration, and in January 2012 with 5 antennas in the extended (A) configuration.
The on-source time of this dataset is $\sim 4.6$ hr in the compact configuration, and 
$\sim5.9$ hr in the extended configuration.

Data reduction was performed using GILDAS.
The system temperatures during the observations were in the range between 150 and 300 K.
The absolute flux
calibration relies on the strong quasars 3C273 and 3C279 and its accuracy is expected to be of the order 10\% (Castro-Carrizo \& Neri 2010).  
The synthesized beams, obtained by using natural weighting,  are $5.6\arcsec\times4.6\arcsec$ for the D configuration 
and   $1.4\arcsec\times0.7\arcsec$ for the A configuration maps.
The achieved
noise levels are 0.9 and 1.2 mJy/beam over 20 MHz  (i.e. $\sim$50 km/s)
for the D configuration data, and configuration A data,
respectively.

\section{X-ray observation and data analysis}

\object{NGC6240} was observed by Chandra on July 2001 for about 35 ksec. 
Reduced and calibrated data are available from the public CXO data
archive. Results from these observations have been published by Komossa
et al. (2003), and Lira et al. (2004).

\section{Results}

\begin{figure}[t]
\centering
\includegraphics[scale=0.6]{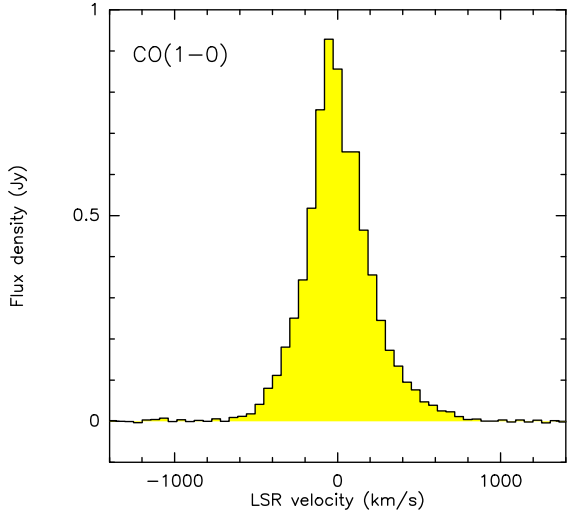}
\caption{\co~  spectrum of \object{NGC6240} obtained with IRAM/PdBI in D configuration. Asymmetric
broad wings extend from $-600$ km/s to $+800$ km/s with respect to the systemic velocity.
The 3 mm continuum has been subtracted in order to highlight the high velocity wings. The spectral channels are 53.3 km/s wide.}
\label{spettro_tot}
\end{figure}

Figure \ref{spettro_tot} shows the continuum-subtracted spectrum of  the \co~
line, extracted from a polygonal region enclosing the source from 
the D array configuration data. 
The 3 mm continuum was estimated by averaging the visibilities in the spectral channels corresponding to the velocity ranges -3500 to -2000 km/s, and 2000 to 4000 km/s with respect to the systemic velocity. This range of velocities (not shown in Fig. 1 for clarity) is fully covered by the WideC Correlator and it is free from emission lines.
Based on the data taken in the D array configuration, the continuum emission peaks at $-0.16, 0.82$ arcsec off the phase tracking center, and has a flux density of 12.7 mJy.  This is consistent with the 1 mm continuum reported by Tacconi et al. (1999), assuming a radio spectral index of $0.7$.  
Two components of the radio and mm continua, centered at the position of each AGN, are found in maps with higher spatial resolution (Tacconi et al. 1999, Colbert et al. 1994). Our data from the D configuration do not allow to spatially resolve these two components.
The continuum map shows one component whose fitted size is  $1\pm0.1 \rm kpc$ (FWHM), assuming a circular gaussian model.  
The 1 mm and radio (8 GHz) continua are consistent with non thermal synchrotron emission.
The CO line peaks close to the assumed systemic velocity ($\sim-50$ km/s).
Broad and asymmetric wings extend to at least
$-600$ km/s on the blue side and $+800$ km/s an the red side of the line peak.
The full width at zero intensity ($\rm FWZI\sim1400$ km/s)
is broader than that of \cod~ and \cot~ reported by Engel
et al. (2010) and Iono et al.  (2007). 
In particular, the blue side of
the line covers a larger velocity range than the previously reported  $-$450 km/s 
(Bryant \& Scoville 1999, Tacconi et al. 1999, Engel et al. 2010, Iono et al. 2007, Nakanishi et al. 2005), probably due to 
the larger bandwidth and the better sensitivity of the new PdBI receivers.

\begin{figure}[t]
\includegraphics[width=9cm]{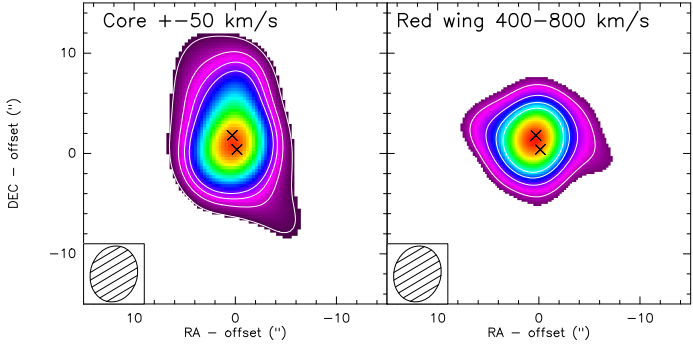}
\caption{ Maps of the CO core emission (left panel, $-50$, +50 km/s) and red wing
emission (400 to 800 km/s with respect to the systemic velocity), from the compact D array data. 
Each contour is 5$\sigma$ (limited to 20$\sigma$). The positions of the two AGN nuclei are shown by crosses.
The synthesized beam is shown in the bottom-left corners.}
\label{core}
\end{figure}

Figure \ref{core} shows the integrated maps from the D configuration data of the CO core emission
( $-50$ to 50 km/s) and of the red-shifted velocities 
(from 400 to 800 km/s with respect to the systemic velocity). 
The axes show the coordinate offsets with respect to the phase tracking center, (ra, dec)$=$(16:52:58.9, 02:24:02.9).
The positions of the two AGN nuclei from VLBI observations (Hagiwara et al. 2011) are indicated by crosses.
The CO core emission is elongated in the north-south direction on scales of 10\arcsec, and shows 
a faint south-western elongation.  
The map of the red wing shows a strong compact source, of size $\sim$1.7 kpc, co-spatial with the narrow core emission, 
and an elongation in the east-west direction with a position angle of 80 degrees.  
Fitting an elliptical gaussian model in the uv plane gives a flux of 11.4 Jy km/s for this high velocity, red-shifted component.
The uv-fit results are reported in Table 1.

\subsection{Blue-shifted CO emission}

\begin{figure*}[t]
\centering
\includegraphics[width=15cm]{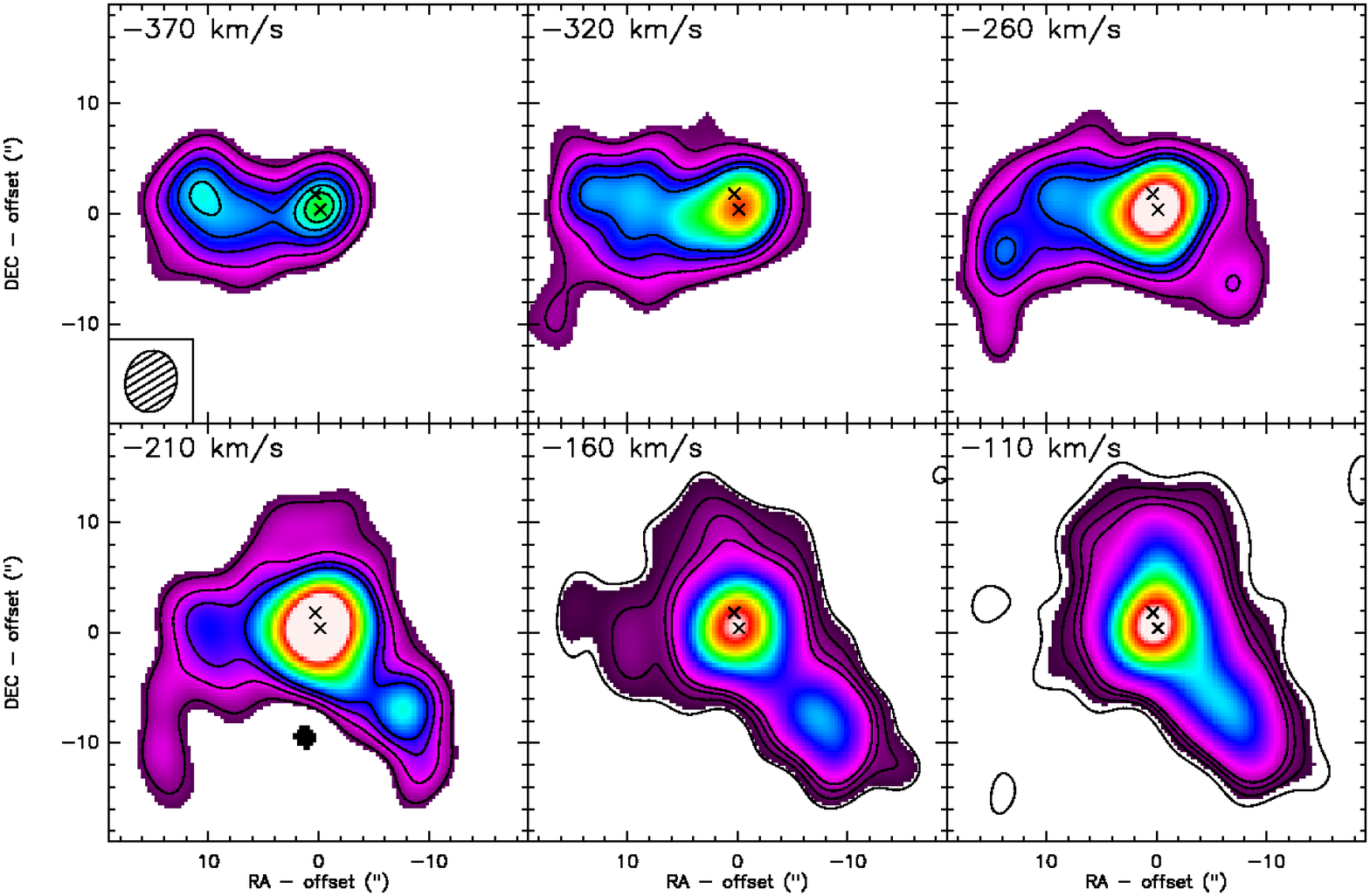}
\includegraphics[width=15cm]{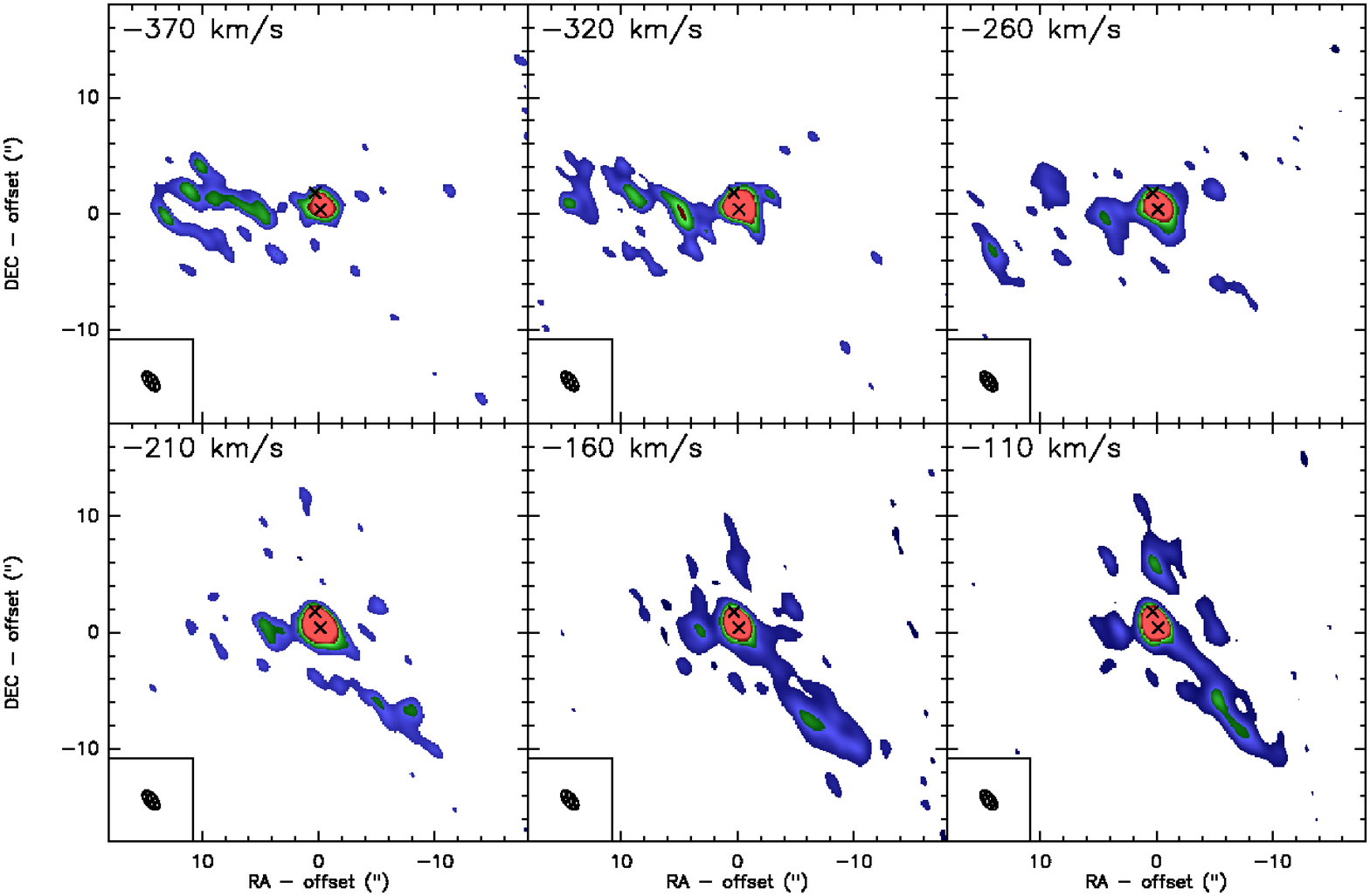}

\caption{Upper panel: \co~ maps from the compact array data, in velocity bins of 20 MHz each (velocity labels are rounded off), showing the detection of blue-shifted CO, including structures extended on scales of 10-15\arcsec. The positions of the two AGN nuclei are shown by crosses. The synthesized beam is shown on the first panel only, for clarity. Each contour is 5$\sigma$ (limited to 20$\sigma$). 
 Lower panel:  maps from merged data of the D and A configurations in the same velocity channels.  The synthesized beams are shown in the bottom-left corners.}
 \label{mapscda}
\end{figure*}

We now examine in detail the blue-shifted emission of CO. 
We find complex morphology, extended on scales from a few arcseconds to 15-20
\arcsec. Figure \ref{mapscda} shows \co~ maps at different velocities,  
from $-400$ to$-100$ km/s in channels 20 MHz ($=$53.3 km/s) wide,  for the
D and  A configuration data. 
Each contour in Fig. 3 is 5$\sigma$ (limited to 20$\sigma$ for clarity). 
Two structures are particularly prominent:  emission extended eastward 
out to at least 15\arcsec~ with velocities from $-400$ to $-200$ km/s, 
and emission extending southwestward with velocities from $-200$ to $-100$ km/s.  
The most prominent emission is located eastward from the nuclei,
 i.e. approximately perpendicular to the line connecting the galactic nuclei,
in the velocity range $-400$ to $-150$ km/s.
From this, a structure showing velocities of  $\sim -260$ km/s develops in the southern direction, 
likely a tidal tail remnant of the merger.  This shows substructure, in the form of three main clumps of CO 
emission, and it coincides with the smooth structure seen by Bush et al. (2008) at 8 $\mu m$, and tracing dust through 
emission by poly-aromaticÊhydrocarbons (PAHs). 
Figure \ref{mapscda} (lower panels) shows the maps obtained by merging the data from the D and A configuration.
The synthesized beam is intermediate between those of the two and allows for better spatial resolution of the thin, jet-like structures, to better follow their alignment with the emission at other wavelengths. 
In the merged maps, we estimate, that for the central region (around the nuclei) we are missing $\sim 33\%$
of the flux.

\begin{figure}[b]
\centering
\includegraphics[width=7cm]{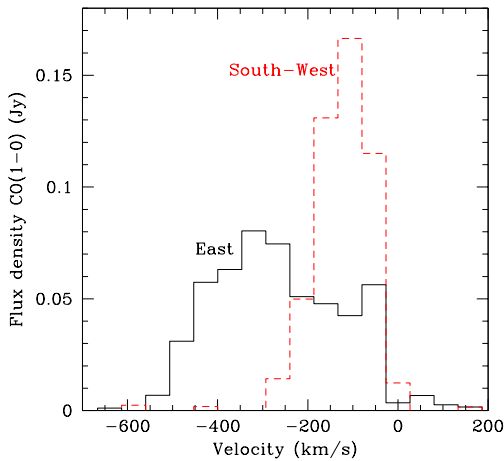}
\caption{\co~ spectra of \object{NGC6240} extracted from the D configuration data in circular regions of
  2\arcsec~ radii encompassing the extended, blue-shifted emission 
  eastward (solid black histogram) and southwestward of the nuclei (red
  dashed histogram). The extraction regions are centered at  offsets of (7.6",1.3")  and  (-6.4",-6.8") from the phase tracking center. }
\label{spettro_blobs}
\end{figure}

Figure \ref{spettro_blobs} shows two spectra
extracted from circular regions of 2\arcsec~ radius centered on the
eastern and southwestern features. 
The spectra were extracted from the cleaned data cubes in regions that enclose the extended structures shown in Fig. 3. 
These spectra are presented for the purpose of showing the emission line peak velocity and line-width, 
and should not be used to derive the fluxes. Here we derive the line fluxes from the visibilities of D configuration data.
We derive the line intensities of the  two blue-shifted structures and of the nuclear region 
by fitting in the uv-plane the visibilities  of the compact D configuration array data. 
The fit in the uv-plane yields the flux at zero-spacing.
First, we fit the central region around the two nuclei (see Fig. 2, left panel),  with the combination of two elliptical gaussian models, which yield a 
line intensity $\rm I_{CO}=213$ Jy km/s. The results of the fit are reported in Table 1. 
The fit produces a residual table where the visibilities of the central component have been subtracted.
To derive the line intensity of the blue-shifted, extended structures, we fit the residual visibilities using two elliptical gaussians. 
The derived line intensity  is 49.3 Jy km/s (over 600 km/s) for the eastern emission region, and 32.5
Jy km/s (over 400 km/s) for the south-western streamer. 
Summing these three components, we obtain a total integrated CO intensity of $295\pm29$ Jy km/s, in agreement with both 
Solomon et al. (1997) single dish observations (310 Jy km/s), and with the interferometric flux (324 Jy km/s ) of  Bryant \& Scoville (1999).

Note that the emission of each blue-shifted region is 4 to 7 times fainter than the central part of the galaxy.
As seen in Fig. 3, both these structures are spatially resolved. 
The eastern component is found $7.6\arcsec,1.3\arcsec$ off the phase tracking center. 
The southwestern one is found at $-6.4\arcsec,-6.8\arcsec$ off  the phase center.
The fit with elliptical gaussians gives sizes of  $14.4\arcsec\times8.2$\arcsec~ for the eastern blob, 
and $8.3\arcsec\times4.4$\arcsec~ for the southwestern blob.

\begin{table*}
\centering
\caption{\label{t1}Measured quantities from the uv-plane fit,  and derived quantities (line luminosities and gas masses) of the CO emitting components.}
\begin{tabular}{lccccccccc}
\hline
Spectral & RA & DEC & S$_{\nu}$ & SIZE  &  PA  & FWZI & $\rm I_{CO}$ & L'(CO) & M(H$_2$) \\
Component & [J2000] & [J2000] & [mJy] & [$\arcsec$] & [deg] & [km/s] &  [Jy km/s]  &   [$10^9~ \rm K~ km~ s^{-1} pc^2$]   &   [$10^9~\rm M_{\odot}$]\\
\hline
\hline
Nuclear   &     $16:52:58.90$  & $02:24:04.16$  &  138.3$\pm$0.8    &    $4.7\times3.4$   &  $0.4$ & 1400 &  $213$  & 5.7 &  4.5  \\
Red wing &     $16:52:58.93$  & $02:24:04.35$  &   28.5$\pm$0.7    &    $3.4\times2.4$   &  $81 $       &400& $11.4$  & 0.3 &  0.15    \\
Blue-E      &     $16:52:59.41$  &  $02:24:04.34$ &   82.1$\pm$5         &   $14.4\times8.2$  &  $-65$      &600&   $49.3$  & 1.3 &  0.7  \\
Blue-SW     &  $16:52:58.47$ &  $02:23:56.15$   &   81.3$\pm$2         &   $8.3\times4.4$   &   $38$    &400&  $32.5$  &  0.87 &  0.43  \\
\hline
\end{tabular}
\tablefoot{Data were fitted with elliptical gaussian models in the uv-plane on the continuum-subtracted visibilities. Errors are of statistical nature and do not account for the uncertainties in the absolute flux calibration. This is conservatively expected to be of the order 10$\%$.}
\end{table*}

\begin{figure*}
\centering
\begin{tabular}{cc}
\includegraphics[width=16cm]{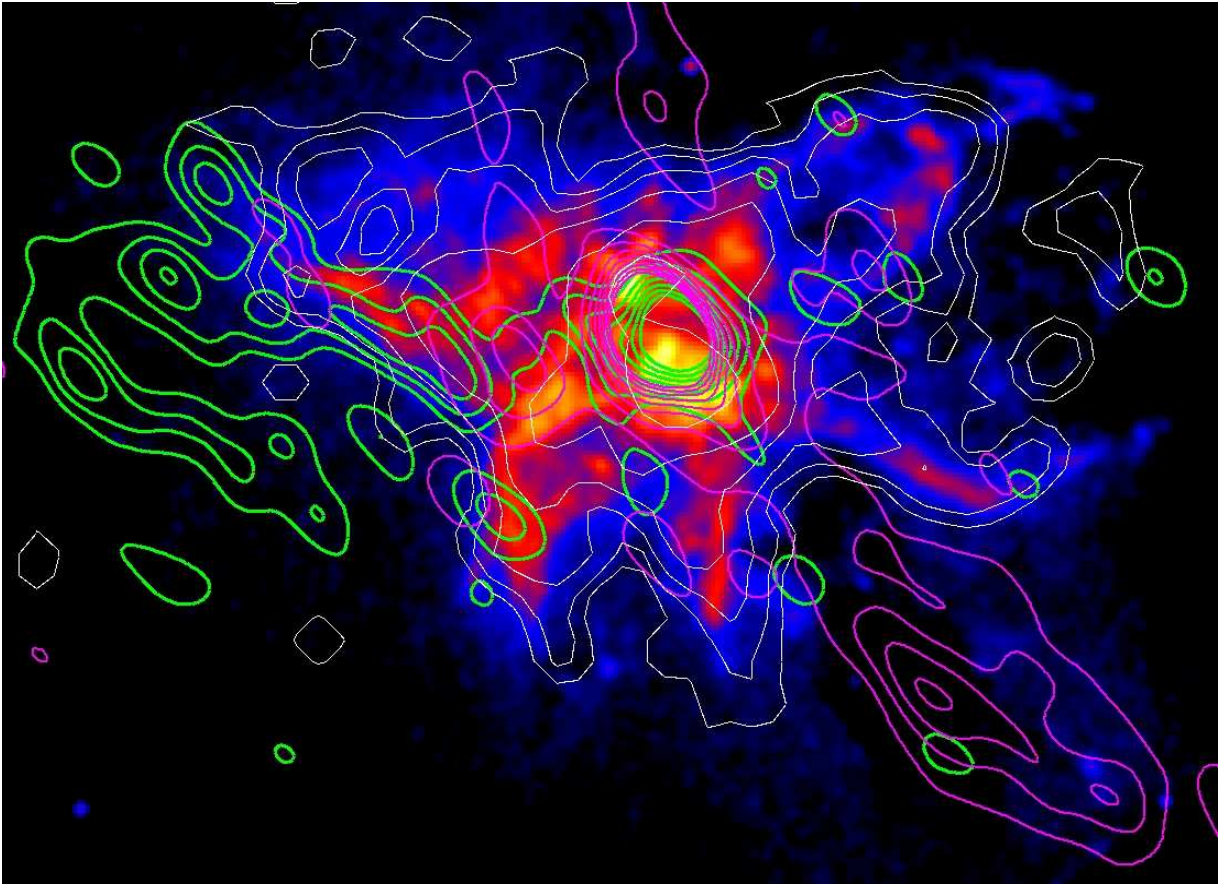}
\end{tabular}
\caption{H$\alpha$ map of \object{NGC6240} (color image). \co~ emission at different
  velocities: $-350$ km/s (green contours), $-100$ km/s (magenta contours),
  with respect to the system velocity. Contours are calculated by merging D and A configuration 
  data. Chandra 1.6-2 keV emission is shown by white contours. 
}
\label{halpha}
\end{figure*}

Figure \ref{halpha} (left panel) shows the Wide Field Planetary Camera (WFPC2 F673N)
image from the Hubble Space Telescope (HST), which includes the
 galaxy's H$\alpha$ emission (Gerssen et al. 2004), with  overlayed contours of the blue-shifted \co~ emission at 
$-400$ km/s and $-100$ km/s. 
The WFPC2 image shows that the  H$\alpha$ nebula comprises five main bright filaments: two southwards of the nuclear region, one located in the western region, and two eastwards from the nuclei.
We note that the CO emission with velocity -100 km/s is located on the
southwestern dust lane, in between two H$\alpha$ filaments.
An elongation of this component toward the northern dust lane is also visible.  
The CO emission centered at  $-400$ km/s  first
follows the eastern elongation of the H$\alpha$ emission, and continues
further eastward and southward.   
We also show 
the X-ray emission at 1.6-2 keV, centered on the highly ionized Si
emission (white contours).  Note that the X-ray data trace remarkably well the
H$\alpha$ emission. In particular, the soft X-ray emission is 
coincident with the eastern H$\alpha$, H$_2$ (see fig. 9 in Max et
al. 2005) and CO elongation. This leads us to investigate further
the association of the X-ray emission with the H$\alpha$, H$_2$ and
CO emitting gas.

\subsection{X-ray spatially resolved spectroscopy}

We extracted a spectrum from the X-ray Chandra data at the position
of each of the  five H$\alpha$ filaments described above, and combined them. 
The extraction regions are shown in Figure 6. 
A background spectrum has
been extracted from a source free region at distances of 2 to 5 arcmin from
the nuclei of the galaxy of 25 arcmin$^2$ size, in order to avoid the contamination 
from the diffuse X-ray emission, which is still seen on scales of 1 arcmin away from the nuclei.
The background-subtracted X-ray spectrum is plotted in
Fig. \ref{xrayspec}. Strong emission lines are visible at about
1.3-1.4 keV, 1.7-1.9 keV and 2.3-2.4 keV. At these energies the ionized
emission from Mg, Si and S is expected. We fitted the spectrum using
{\sc XSPEC} and adopting $\chi^2$ statistics. The
spectrum was binned to have at least 30 counts per channel.  We
limited the fit to the 0.5-7 keV band (440 original channels, 67
bins), where the instrument response is best calibrated and to avoid
strong background lines at high energy (see e.g. the Chandra
background spectrum in Fiore et al. 2011). We started modeling the
spectrum with a thermal equilibrium gas component ({\sc MEKAL} in {\sc
  XSPEC}), reduced at low energy by photoelectric absorption by gas
along the line of sight. This model is clearly inadequate to reproduce
the observed spectrum, giving a $\chi^2$=179.5 for 63 degrees of
freedom (DOF) and large residuals at 0.8-0.9 keV, 1.2 keV, 1.8 keV and
above 4 keV. We then added a second thermal equilibrium gas component
to the model.  Figure  \ref{xrayspec} shows the best fit model with 8
free parameters (two temperatures, two metal abundances, two
normalizations and two absorbing column densities).  The best fit $\chi^2$
is 93.3 (59 DOF).  Note the rather strong positive residuals at 1, 1.4,
1.8, 2.2 keV, i.e. the position of the Mg, Si, and S line
complexes. Stronger line emission 
is expected in non-equilibrium models, because of the broader ion
distribution with respect to thermal equilibrium models at the same temperature and metal abundances.  
In particular, shock models, like {\sc XSPEC PSHOCK} are known to produce
spectra with prominent line emission. We therefore fitted the spectrum
with a model including a thermal equilibrium component and a shock
component. The best fit $\chi^2$ is now 62.1 (58 DOF).  The
improvement in $\chi^2$ with respect to the two component thermal
equilibrium model is significant at the 99.9997\% confidence level
(using the F test). Residuals with respect to the best fit model do
not show any systematic deviation. Figure \ref{eufspec} shows the best
fit unfolded (i.e. corrected for the response matrix of the instrument)
spectrum with the identified contributions of the thermal
equilibrium and shock components.  We conclude that the X-ray analysis
supports the idea that shocked gas is present at the position of
strong H$\alpha$ emission, both in the nuclear starburst and in the
elongated filaments. 

\begin{figure}
\centering
\includegraphics[width=7cm]{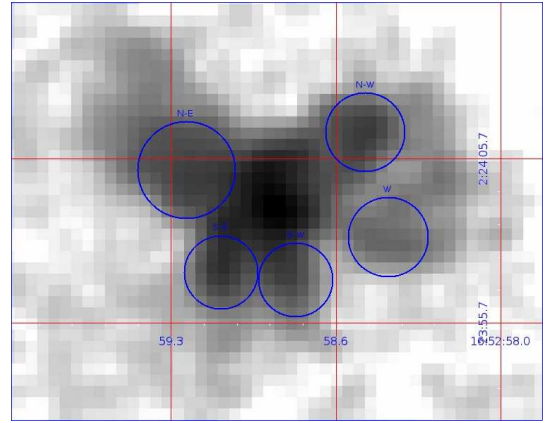}
\caption{The Chandra  X-ray map of NGC 6240 in the energy range 0.3-4 keV. 
The circles indicate the regions were we extracted the X-ray spectra. The region 
for the background extraction is outside the limits of this map, at 2-5 arcmin from the nuclei.}
\label{xrayreg}
\end{figure}

\begin{figure}[!]
\centering
\includegraphics[width=6.5cm,angle=-90]{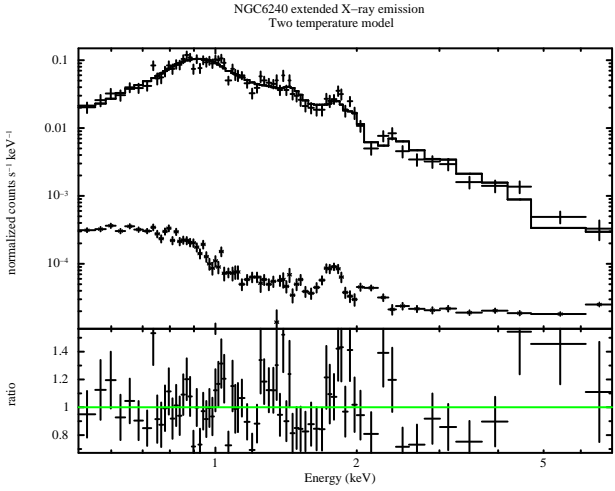}
\caption{Top panel: the top curve represents the background corrected Chandra spectrum (uncorrected for the response of the instrument;
ÊÊÊÊÊÊÊcrosses denote spectral widths and amplitude uncertainties) extracted at the position of H$\alpha$
  filaments (see Fig. 6), and fitted with a two-temperature 
  thermal equilibrium model (solid line). The lower curve represents the background spectrum.  
  Data, represented by crosses,  have been binned to give a
  signal to noise ratio $>5$ in each bin (for plotting purposes only). 
  Bottom panel:  the data to model ratio (the green solid line representing a ratio of unity).  Note the 
  residuals at about 1 keV, 1.4 keV, 1.8 keV, 2.2 keV and 5 keV. }
\label{xrayspec}
\end{figure}

\begin{figure}[!]
\centering
\includegraphics[width=7.0cm,angle=-90]{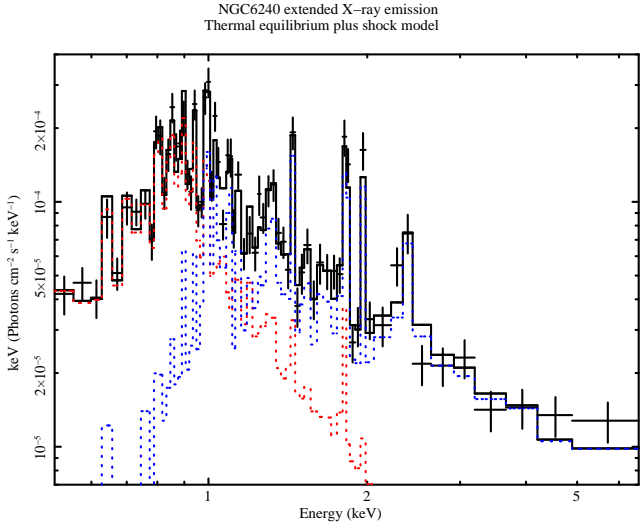}
\caption{The  background corrected Chandra spectrum (also corrected for the response matrix of the instrument) extracted at the position of the 
  H$\alpha$ filaments (see Fig. 6). Symbols are as in Fig. 7.  Red dotted
  histogram: thermal equilibrium model component;  blue dashed histogram:  shock model component.} 
\label{eufspec}
\end{figure}

\section{Discussion and conclusions}

We have obtained deep 3 mm maps of the archetypical interacting galaxy
\object{NGC6240} with the IRAM PdBI, covering a velocity range of 10.000 km/s.  
The CO(1-0) line shows strong blue and
red wings extending  from $-600$ km/s to +800 km/s with
respect to the systemic velocity.  The line (FWZI$=1400$ km/s) is
significantly broader than that previously reported by Tacconi et al. (1999), Engel et al. (2010), and 
Iono et al. (2007).  
The systemic CO emission shows a north-south elongation over at least 10$\arcsec.$
Elongation in the same direction is seen in the  \co~ maps of Bryant \& Scoville (1999), and in the 
\cod~ maps of Tacconi et al. (1999), and  Engel et al. (2010) on 10 times smaller scales.
The CO luminosity of  this region is $\rm L'(CO) = 5.7~10^9~ K~ km~ s^{-1}~ pc^2$.
We derive an estimate of the molecular gas mass in this region, assuming the standard CO to $\rm H_2$ conversion factor,  
$\rm \alpha=0.8 ~M_\odot$ (K km s$^{-1}$ pc$^2$ )$^{-1}$ (units omitted hereafter). 
We find $\rm M(H_2) = 4.5~ 10^9~ M_{\odot}$, consistent with the value derived from CO(2-1) for this region by Tacconi et al. (1999).

We were able to identify new components.
We find CO emission extended up to distances of 15-20\arcsec~ from the galaxy
centers (7-10 kpc at the distance of \object{NGC6240}). In particular, we find
strong emission blue-shifted by $\sim 150$  to 400 km/s extending eastward by
at least 15\arcsec~ from the nuclei, and by
100-200 km/s extending south-westward on a similar scale. 
The presence of the latter component was suggested by the interferometric  CO maps of 
Bryant \& Scoville (1999), although with low significance.

The CO southwestern emission coincides with the dust
lane seen in HST images (Gerssen 2004, also see Fig. \ref{halpha}) and in the IRAC 8 $\mu$m image (Bush et al. 2008). 
This large scale dust distribution has been interpreted as due to a tidal
tail curving in front of the system (Gerssen et al. 2004, Yun \& Hibbard 2001). 
Molecular gas is 
associated with this tidal tail, a situation reminiscent of M82, where
Walter et al.  (2002) found molecular gas in the tidal tales correlated
with dust absorption features. 
The integrated CO luminosity of
the southwestern emitting region is $\rm L'(CO)=8.7\times10^8$ K km s$^{-1}$ pc$^2$.
For the conversion from  CO luminosity into molecular gas mass M(H$_2$),
we conservatively adopted the lowest conversion factor found in the giant outflows 
and streamers of M82 (Weiss et al. 2001),  $\rm \alpha=0.5$ . 
The mass of the molecular gas in this region 
is thus $\rm M(H_2)=4.3\times10^8~ M_\odot$,  4-to-10 times
the H$_2$ mass in the streamers of M82 (Walter et al. 2002, derived for the same conversion factor). 
This estimate likely represents a lower limit to the molecular gas mass in this streamer. 
The physical size of the southwestern tidal tail is however at least 15
\arcsec, i.e. 7 kpc, 4-7 times larger than the streamers in M82 (Walter
et al. 2002). 
The CO extended emission to the
north (see Fig.2, left panel)  coincides with a dust lane seen in HST images (Gerssen et al. 2004), and might be
associated with another streamer.
The detection of molecular
streamer(s) in \object{NGC6240} confirms that the
molecular gas is severely affected by galaxy interaction, and that the
redistribution of molecular gas is likely the trigger for the strong
starburst activity in the central region of \object{NGC6240}.

The blueshifted eastern CO emitting region is not associated with the dust lanes mentioned above,
but follows  a H$\alpha$ filament, and PAH emission observed at 8 $\mu$m  (Bush et al. 2008). 
The emission-line
nebula seen in H$\alpha$ images (Gerssen et al. 2004) is 
interpreted as evidence of a superwind that is shock heating ambient
ISM. The H$\alpha$ emitting filaments are aligned east-westward, perpendicular
to the dust lanes and to the line connecting the two nuclei. 
The X-ray emission is associated with the H$\alpha$ filaments (Lira et
al. 2002). In particular, strong soft X-ray emission is coincident
with the north-eastern filament (N-E region in Fig. 6). We re-analyzed the Chandra X-ray data and found
strong evidence for shocked gas at the position of the H$\alpha$
filaments. The presence of shocked gas in the \object{NGC6240} system is
confirmed by the detection of strong H$_2$(1-0) S(1) emission at
2.12 $\mu$m. \object{NGC6240} has the strongest H$_2$ line emission found in any
galaxy (Goldader et al. 1995). Shocks are usually identified as
the excitation mechanism for this line. 
Ohyama et al. (2000, 2003)
suggest that such shocks are produced by a superwind outflowing from
the southern nucleus and colliding with the surrounding molecular gas.
Intriguingly, Max et al. (2005) find that excited H$_2$
emission closely follows the H$\alpha$ filament extending eastward.
Note that the position of one of the peaks of the eastern blue-shifted
CO emission coincides with the position where Bland-Hawthorne et
al. (1991) and Gerssen et al. (2004) found a large velocity gradient of the
ionized gas (the velocity decreasing, roughly, north to south). 
No significant radio continuum emission is detected in this region in the VLA maps of Colbert et al. (1994).
The blue-shifted molecular gas in this region might be a tidal tail, left behind during the merger. 
However, its association with strongly shocked gas
suggests that a shock is propagating eastward and is compressing also 
the molecular gas, while crossing it.

The integrated 
luminosity of CO in this region is $\rm L'(CO)=1.3\times10^9$ K km s$^{-1}$ pc$^2$, corresponding to 
a gas mass M(H$_2$)=$7\times10^8$ M$_\odot$ 
(assuming again  $\alpha$ = 0.5). 
Hypothesizing that this outflow originates from the southern, luminous AGN (or from both the AGN), we derive here a mass loss rate. 
We assume that the molecular outflow is distributed in a spherical volume of radius $\rm R_{of}= 7~ kpc$  (i.e. the distance of the eastern blob from the AGN), 
centered on the AGN.  
If the gas is uniformly distributed in this volume, the volume-averaged density of molecular gas is given by 
$<\rho_{of}> = 3~ M_{of}/\Omega ~ R^3_{of}$, where $\Omega$ is the solid angle subtended by the outflow and $\rm M_{of}$ is the mass of gas in the outflow (Feruglio et al. 2010, Maiolino et al. 2012). 
This assumption is an approximation, and evidently cannot represent the complexity of the system, but can provide a rough estimate 
of the mass loss rate.   
Based on this geometry, we can derive a mass loss rate by using the relation: 
\smallskip

$\rm  \dot M(H_2)  \sim v~ \Omega~ R^2_{of}~ <\rho_{of}> = 3~ v ~\frac{M_{of}}{R_{of}}$

\smallskip
\noindent
where $v$ is the terminal velocity of the outflow ($\sim 400$ km/s).
This relation yields a mass loss rate of $\rm \dot M \sim 120 ~M_\odot/yr$. 
The  H$\alpha$, H$_2$ and \co~maps suggest that the outflow is most likely conical. 
In this geometry, since the mass loss rate is independent of $\Omega$,  the mass outflow rate would be equal to the spherical case if 
there are no significant losses through the lateral sides of the cone.  
As it is observed in other local massive outflows (Cicone et al. 2012, Aalto et al. 2012),  the outflowing gas is likely characterized by  a 
large range of densities, ranging from low density gas to dense clumps, which would increase the mass flow rate. 
In addition, the mass flow rate would obviously be larger than our previous estimates if $\alpha$ is significantly
higher than 0.5. To date, this is the lowest conversion factor measured in an extragalactic object.
This said, it is unlikely that the mass loss rate is smaller
than several tens $\rm M_\odot yr^{-1} $, and it is likely as big as a few
hundreds $\rm M_\odot yr^{-1}$.  
The kinetic power of the outflowing gas is given by the relation $\rm P_k = 0.5~ v^2~ \dot M = 6~ 10^{42}~ erg~ s^{-1}$.
The age of the outflow is  $>2\times10^7$ years, since it is observed at about 7 kpc distance from the southern nucleus.
The star-formation rate at the position of the
eastern filament can be evaluated through both the H$\alpha$ luminosity
and the X-ray luminosity (e.g. Kennicut 1998, Ranalli et al. 2003).
The 0.5-2 keV X-ray luminosity at the position of the East filament is
$\rm 0.5-1\times10^{41}~ erg~s^{-1}$, which according to Ranalli et al. (2003)
would imply a star-formation rate of 10-20 $\rm M_\odot yr^{-1}$. 
This estimate is derived by assuming that all the X-ray luminosity is due to star-formation. 
In section 4.2 we showed that at least a fraction of the X-luminosity is due to a shock, therefore the 
SFR derived from the X-ray is an upper limit. 
This would suggest that the outflow is not pushed by SN winds. 
Indeed, the power transferred to the ISM by a star-formation driven wind is given by 
$\rm P_{SF} = \eta \times  7\times 10^{41} \times SFR =10^{42}~ erg~s^{-1}$, 
where $\eta \sim 0.1$ is the standard mass-energy conversion (see e.g. Lapi et al. 2005).
We conclude that it is unlikely that the molecular flow is powered by star-formation. 
Instead, star-formation in this region is likely in the process of being quenched by the outflow. 
However, we cannot exclude that star-formation in this area is induced by the compression 
caused by the propagating shock.

We detected a red-shifted component with velocity 400 to 800 km/s with respect to the 
systemic velocity (Fig. 2), centered around the two AGN nuclei. 
Interestingly this emitting region is elongated in the same east-west direction as the blue-shifted emission discussed above,
although on smaller scales ($\sim 1.7$ kpc in diameter).
We derive a CO luminosity of this component of L'(CO)$=3.0\times10^{8}$ K km s$^{-1}$ pc$^2$, 
which converts into a gas mass of M(H$_2$)$=1.5\times 10^{8}$ M$_{\odot}$, under the same assumptions given above for 
the CO-to-H$_2$ conversion factor.
The large velocity of this component suggests that the AGN might contribute to the dynamics of this gas (Sturm et al. 2011).
 
Given the complex dynamics and morphology of this system, it is not trivial to disentangle and quantify the 
relative role of each mechanism. 
Probably several mechanisms are acting 
contiguously: mainly the radiation pressure of the AGN together with 
dynamic shocks induced by the merger event.
High resolution, X-ray observations will help to clarify the interaction between the star-forming regions 
and the CO extended structures (Wang et al. in preparation, conference communication).
The high spatial resolution data taken in the A array configuration indeed provide new insights on the nuclear region, 
which will be addressed in a separate publication (Feruglio et al. 2012, in preparation).

\begin{acknowledgements}
We thank D. Downes and R. Neri for useful inputs and careful reading of the paper. F.F. acknowledges support from PRIN-INAF 2011.
We acknowledge the referee for her/his very careful review, that allowed us to significantly improve the quality of this work.
\end{acknowledgements}

\end{document}